\documentclass{article}
\usepackage[pctex32]{graphics}
\title{New Spin-Wave Mode in Weak Ferromagnetic Fermi Liquids}  
\author{Penka I. Petkova\thanks{I thank Prof. K. S. Bedell for suggesting the project and for valuable discussions during the early stages of this paper. I also thank K. Blagoev, P. Gopikrishnan, J. Engelbrecht and R. McQeeney for useful communications. I acknowledge support from the Department of Physics, Boston College, where most of the work was done. This work was sponsored in part by DOE Grant $DEFG0297ER45636$.; Email: petkova@ingress.com} \\ Center for Advanced Technologies, P.O. Box 680, New York, NY 10159}
\date{(November 22, 1999)}
\begin{document}
\maketitle
\begin{abstract}
We study a phenomenological model for weak ferromagnetic Fermi liquids and investigate the properties of the spin waves in the model. The Landau kinetic equation is used to derive, in addition to the known Goldstone mode, a new spin-wave mode -- the first Silin-like ferromagnetic mode. We discuss the role of the interaction parameter $F^a_1$ on the behavior of the Goldstone mode and the first Silin-like ferromagnetic mode. 
\end{abstract}

Landau introduced the Fermi liquid theory in 1956 \cite{Ref1} to describe the low-temperature 
thermodynamic properties, transport and collective phenomena in strongly correlated Fermi systems. The theory of Fermi liquids was used to consider spin waves in ferromagnetic metals \cite{Ref2} and further developed into the framework of the ferromagnetic Fermi liquid theory (FFLT). The FFLT can be used to study weak ferromagnetic metals such as $MnSi$, $ZrZn_2$, $Ni_3Al$, subject to increasing interest for their physical properties and potential applications. Many attempts to understand these systems focus on the dynamics of the magnetic excitations, a complex problem, both theoretically and experimentally. A direct experimental information on magnetic excitations in ferromagnetic metals can be obtained from inelastic neutron scattering measurements. In weak magnets the scattering intensity is relatively low, which makes the measurements difficult. The experimental results for $Ni_3Al$ are limited \cite{Ref3}, but there are almost complete measurements for $MnSi$ \cite{Ref4}.  

In this paper we derive the dispersion relations for spin waves in weak ferromagnetic Fermi liquids (at $T=0$ $K$), using the FFLT, with the following main results: (i) Two modes are obtained -- the well known Goldstone mode and a new mode, similar to the first Silin mode in the case of para magnets in a magnetic field \cite{Ref5}, which we call first Silin-like ferromagnetic mode; (ii) In the hydrodynamic limit (small $q$-limit, where $q=|\vec{q}|$ is the momentum transfer) our model leads to the usual quadratic dispersion law $\omega(q)\sim q^2$ for the Goldstone mode, but there are deviations from the hydrodynamic results for the Goldstone mode and for the first Silin-like mode for larger values of $q$; (iii) The Goldstone mode touches the Stoner excitation boundary tangentially, as predicted by the random phase approximation (RPA) theory \cite{Ref6}, but in contrast with the RPA our model gives different shape for the $\omega(q)$ curve in the region close to the particle-hole continuum.

To study the collective excitations in a ferromagnetic metal we investigate the time evolution of the quasi-particle distribution function, given by $2\times 2$ matrix in spin space $[n_p(\vec{r}, t)]_{\alpha\beta}=n_p(\vec{r}, t)\delta_{\alpha\beta}+m_p(\vec{r}, t)\sigma_{\alpha\beta}$, where $n_p$ is the quasi-particle density and $m_p$ is the magnetization density. We start with the quasi-classical kinetic equation

\begin{equation}
\frac{\partial n_p}{\partial t}=[n_p, \varepsilon_p]_{P.B.}+[m_p, h_p]_{P.B.}
\end{equation}
where $P.B.$ stands for Poisson brackets. The evolution of the quasi-particle distribution function is generated by the quasi-particle Hamiltonian $[\varepsilon_p(\vec{r}, t)]_{\alpha\beta}=
\varepsilon_p(\vec{r}, t)\delta_{\alpha\beta}+h_p(\vec{r}, t)\sigma_{\alpha\beta}$, where  $\varepsilon_p$ is a mean quasi-particle energy and $h_p=-B+2\sum_{p\prime}f^a_{pp\prime}m_{p\prime}$ is an effective magnetic field, which includes both coupling to an external field $B$ and to an internal field, due to the quasi-particle interactions (we set the nuclear magnetic moment $\gamma\hbar/2=1$). The interaction parameter $f^a_{pp\prime}$ is the antisymmetric component of the quasi-particle interaction $f^{\sigma\sigma\prime}_{pp\prime}$ (in the limit of weak ferromagnetism $f^{\sigma\sigma\prime}_{pp\prime}$ can be treated as rotationally invariant in spin space \cite{Ref7}).   From Eq.(1) we get a set of coupled equations, describing the time evolution of the quasi-particle density and the magnetization density. For small fluctuations $\delta n_p$ and $\delta m_p$ around the ground state these equations decouple. We are interested only in the equation, which describes oscillations of the transverse components of the total magnetization. The linearized kinetic equation for $\delta m_p$ is \cite{Ref5,Ref8,Ref9}

\begin{equation}
\frac{\partial\delta m_p}{\partial t}+ \vec{v_p}\cdot\vec{\nabla}\left(\delta m_p- \frac{\partial n^0_p}
{\partial\epsilon^0_p}\delta h_p\right)=-2(m^0_p\times\delta h_p+\delta m_p\times h^0_p)
+I[m_p]
\end{equation}
where $I[m_p]$ is  the collision integral, $\delta h_p$ is the fluctuation of the effective 
magnetic field and $h^0_p=\sum_{p^{\prime}}f^a_{pp^{\prime}}m^0_{p^{\prime}}$ is the equilibrium field  due only to the quasi-particle interactions.

We study the response of the system to a transverse perturbation $\delta\vec{B}=\delta B_x\hat i+\delta B_y\hat j$ (the transverse spin polarization $\delta m^+_p=\delta m^x_p+i\delta m^y_p$, there is an analogous equation for $\delta m^-_p$). For excitations close to the Fermi surface we introduce the quantities $\delta m^+_p=-(\partial n^0_p/\partial\epsilon^0_p)\nu^+_p$ and 
$m^0_p=-(\partial n^0_p/\partial\epsilon^0_p)m^0/N(0)$, where $N(0)=k_Fm^*/\pi^2$ is the density of states at the Fermi surface and $m^*$ is the effective mass of the quasi-particle \cite{Ref8,Ref9}. After Fourier transformation of the linearized kinetic equation, Eq.(2), we obtain

\begin{equation}
(\omega-\vec{v}_p\cdot\vec{q}+2m^0f^a_0)\nu^+_p-
\left(2m^0+N(0)\vec{v}_p\cdot\vec{q}\right)\int\frac{d\Omega^{\prime}}{4\pi}f^a_{pp^{\prime}}\nu^{+}_{p^{\prime}}=
-\left(\frac{2m^0}{N(0)}+\vec{v}_p\cdot\vec{q}\right)\delta B^++iI[\nu^+_p].
\end{equation}
We expand both $\nu^+_p$ and the Landau parameters $f^a_{pp^{\prime}}$ in a series of Legendre polynomials, $\nu^+_p=\sum_{\ell}\nu^+_{\ell}P_{\ell}(\hat p\cdot\hat q)$ and $f^a_{pp^{\prime}}=\sum_{\ell}f^a_{\ell}P_{\ell}(\hat p\cdot\hat p^{\prime})$. For the collision term $I[\nu^+_p]$ we use a relaxation-time approximation $I[(\nu^+_p)_{\ell}]=-\frac{1+F^a_{\ell}/(2\ell +1)}{\tau_D}(\nu^+_p)_{\ell}$, where $\tau_D$ is the spin-diffusion lifetime and $I[(\nu^+_p)_{\ell=0}]=0$ since quasiparticle spin is conserved in collisions \cite{Ref8}.

We analyze the possible modes of oscillation using two different approaches: (i) Low-order distortions approximation (hydrodynamic model); (ii) All-order distortions model.

We start with the hydrodynamic model and project out the $\ell=0$, 1, 2 components of Eq.(3). The $\vec{v}_p\cdot\vec{q}$ term in Eq.(3) couples the distortions of order $\ell$ to distortions of order $\ell\pm 1$ (for excitations near the Fermi surface $\vec{v_p}\cdot\vec{q}\rightarrow v_Fq\cos(\hat p\cdot\hat q)$). 

\begin{equation}
\omega\nu^+_0-\frac{1}{3}\nu^+_1\left(1+\frac{F^a_1}{3}\right)qv_F=-\frac{2m^0}{N(0)}\delta B^+  
\end{equation} 

\begin{equation}
\left[\omega+\frac{2m^0}{N(0)}\left(F^a_0-\frac{F^a_1}{3}\right)+\frac{i(1+F^a_1/3)}{\tau_D}\right]\nu^+_1-\nu^+_0(1+F^a_0)qv_F-
\frac{2}{5}\nu^+_2\left(1+\frac{F^a_2}{5}\right)qv_F=
-\delta B^+ qv_F 
\end{equation}

\begin{equation}
\left[\omega+\frac{2m^0}{N(0)}\left(F^a_0-\frac{F^a_2}{5}\right)+\frac{i(1+F^a_2/5)}{\tau_D}\right]\nu^+_2-\frac{2}{3}
\nu^+_1\left(1+\frac{F^a_1}{3}\right)qv_F-\frac{3}{7}\nu^+_3\left(1+\frac{F^a_3}{7}\right)qv_F=0
\end{equation}
where $F^a_{\ell}$ is a dimensionless Landau parameter, defined as $F^a_{\ell}=N(0)f^a_{\ell}$. We take the limit $\delta B^+\rightarrow 0$ and for simplicity work in the extreme collisionless regime ($T=0$ $K$). The distortions $\nu^+_3$ can be neglected in the small $q$-limit if $qv_F\ll\left|\omega_{\ell}(q)-\Delta\right|$, where $\omega_{\ell}(q)$ is the dispersion relation for the spin-wave mode considered in any particular case and $\Delta=-2m^0F^a_0/N(0)$ is the Stoner gap parameter. From Eq.(4), Eq.(5) and Eq.(6) we obtain the dispersion equation 

\begin{equation}
\omega\left[(\omega-\omega^+_1)(\omega-\omega^+_2)-
\left(\frac{9}{5}+F^a_0+\frac{4}{25}F^a_2\right)\left(1+\frac{F^a_1}{3}\right)\frac{(qv_F)^2}{3}\right]+\omega^+_2(1+F^a_0)\left(1+\frac{F^a_1}{3}\right)\frac{(qv_F)^2}{3}=0
\end{equation}
where $\omega^+_1=\Delta(1-F^a_1/3F^a_0)$ and $\omega^+_2=\Delta(1-F^a_2/5F^a_0)$ are the eigenfrequencies of the spin-wave modes $\omega^+_{\ell}(q)$. In what follows we assume $F^a_{\ell}=0$ for $\ell>1$. Then the first solution of Eq.(7) is

\begin{equation}
\omega^+_0(q)=-\frac{(1+F^a_0)(1+F^a_1/3)}{\omega^+_1}\frac{(qv_F)^2}{3}+{\cal O}\left((qv_F)^4\right)
\end{equation}
which we identify as the well known quadratic dispersion law for the $\ell=0$ Goldstone mode. In the limit of weak ferromagnetism $F^a_0\rightarrow -1^-$ and the Goldstone mode is almost independent of the interaction parameter $F^a_1$. The second solution of Eq.(7) 

\begin{equation}
\omega^+_1(q)=\omega^+_1+\left[\frac{(1+F^a_0)(1+F^a_1/3)}{\omega^+_1}-
\frac{4}{5}\frac{(1+F^a_1/3)}
{\Delta F^a_1/3F^a_0}\right]\frac{(qv_F)^2}{3}
\end{equation}
gives the dispersion relation for a new spin-wave mode, the $\ell=1$ mode, which we call first Silin-like ferromagnetic mode. The second term of order $(qv_F)^2$ in Eq.(9) comes in because the distortions $\nu^+_2$ are taken into account; for a certain range of values for the interaction parameter $F^a_1$ this term is comparable in size with the first quadratic term, caused by the $\nu^+_0$ distortions. The assumption $F^a_2=0$ forces the eigenfrequency $\omega^+_2$ of the second Silin-like ferromagnetic mode $\omega^+_2(q)$ to the Stoner gap ($\omega^+_2=\Delta$) and makes the mode strongly damped. The second Silin-like ferromagnetic mode is propagating only if $F^a_2\not=0$.

Eq.(4) and Eq.(5) are the two hydrodynamic equations for the magnetization density -- the net magnetization conservation law and the equation of motion for the magnetization current. The equation for the magnetization current contains a precession term from the internal field $\nu^+_1[2m^0/N(0)qv_F]F^a_1/3$, found also in the paramagnetic case \cite{Ref10}. The precession term in Eq.(5) is missing in the ordinary hydrodynamic equations for ferro magnets \cite{Ref11}; these equations lead to the quadratic dispersion law for the Goldstone mode, and miss completely the first Silin-like ferromagnetic mode. In addition, the equation for the magnetization current contains distortions of the Fermi surface of order $\ell=2$, $\nu^+_2$. While these distortions produce corrections of higher order in the dispersion relation for the Goldstone mode, they are important for the first Silin-like ferromagnetic mode, as shown above.

At present we do not have data for the value of the interaction parameter $F^a_1$ in weak ferro magnets,  
but in other strongly correlated Fermi systems like liquid $^3He$ and $^3He-^4He$ mixtures the range of $F^a_1$ is $|F^a_1|\leq 1$, \cite{Ref9}. We plot in Fig.1 the dispersion of the Goldstone mode, Eq.(8) and the dispersion of the first Silin-like ferromagnetic mode, Eq.(9) for $F^a_0=-1.1$ and for different values of $F^a_1$, $|F^a_1|\leq 1.5$. Changing the value of $F^a_1$ does not influence the behavior of the Goldstone mode, but produces important effects on the first Silin-like ferromagnetic mode. If $F^a_1=0$, the first Silin-like ferromagnetic mode starts exactly from the Stoner gap. The mode is strongly damped because it stays right on the particle-hole continuum, defined as $\Delta\pm qv_F$. If $F^a_1\not=0$, the first Silin-like ferromagnetic mode moves up or down with respect to the Stoner gap, depending on the sign of $F^a_1$ (see Fig.1), and can propagate before entering into the particle-hole continuum and being damped.

\begin{figure}[h]
\includegraphics*[0.2cm,0.2cm][7.13cm,6.2cm]{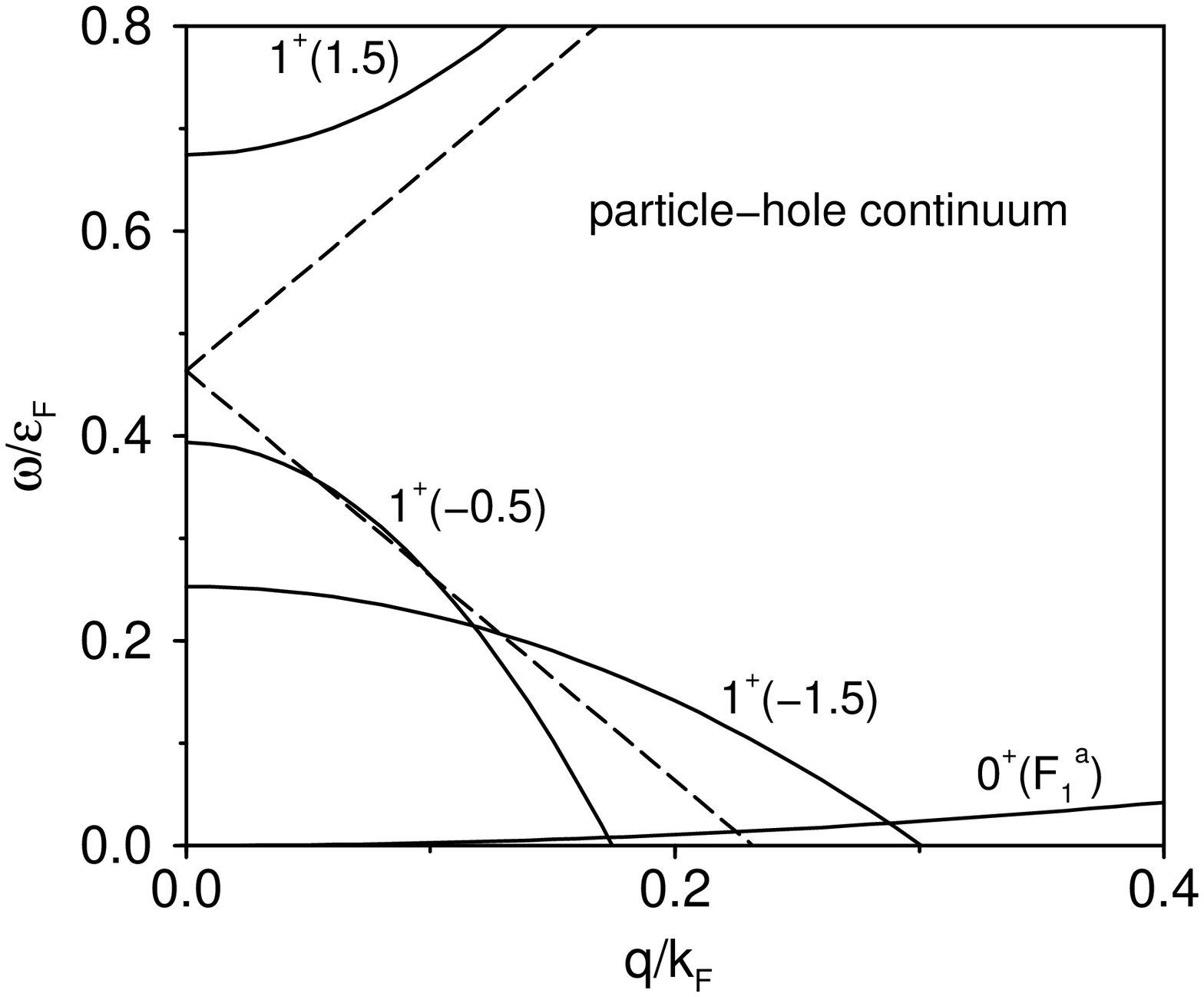}
\caption{The hydrodynamic modes $\omega^+_0(q)\equiv 0^+$ and $\omega^+_1(q)\equiv 1^+$ for ferromagnetic Fermi liquid for $F^a_0=-1.1$ and $F^a_1=-1.5, -0.5, 0, 1.5$ (shown in the brackets).}
\end{figure}

We make a rough estimate of the coefficient in front of $q^2$ in the dispersion relation for the 
Goldstone mode, Eq.(8), and compare it with the coefficient in front of $q^2$ in the empirical dispersion law for spin-wave excitation for $MnSi$, $\hbar\omega_q[meV]=0.13+0.52q^2[\AA^{-2}]$ \cite{Ref4}. We set $F^a_0=-1.1$, $F^a_1=-1.5$, $T_F=300$ $K$, and use the data available for $MnSi$: the equilibrium magnetization under a pressure $P=6.6$ $kbar$ is $m^0=0.389$ $\mu_B/Mn$ \cite{Ref12} and the density of states $N(0)=19$ $states/Ry$ $Mn$ $atom$ $spin$, obtained from band structure calculations \cite{Ref13}. Then Eq.(8) has the form $\omega^+_0(q)=2.5q^2$, where $\omega^+_0(q)$ is scaled with respect to the Fermi energy $\varepsilon_F$ and $q$ -- with respect to the Fermi momentum $k_F$. The empirical law expressed in the same units is $\omega(q)=2q^2$, thus we obtain the same order of magnitude for both  coefficients. 

To analyze the spin-wave modes beyond the small $q$-limit, we go back to the kinetic equation, Eq.(3),  and derive the all order distortions model. As before we expand $\nu^+_p$ and $f^a_{pp^{\prime}}$ in a series of Legendre polynomials and obtain a general expression for distortions of order $\ell$

\begin{equation}
\frac{\nu^+_{\ell}}{2\ell+1} +\sum_{\ell^{\prime}}\frac{F^a_{\ell^{\prime}}}{2\ell^{\prime}+1}\nu^+_{\ell^{\prime}}
\left[\Omega_{\ell\ell^{\prime}}(s)+
\frac{m^0}{N(0)qv_F}\int_{-1}^{1}\frac{P_{\ell}(x)P_{\ell^{\prime}}(x)dx}
{x-s}\right]=\delta B^+\left[\Omega_{\ell 0}+
\frac{m^0}{N(0)qv_F}\int_{-1}^{1}\frac{P_{\ell}(x)dx}{x-s}\right]
\end{equation}
where $x=\cos(\hat p\cdot\hat q)$, $s=(\omega-\Delta)/qv_F$, $\Omega_{\ell\ell^{\prime}}(s)=\Omega_{\ell^{\prime}\ell}(s)=\frac{1}{2}\int_{-1}^{1} \frac{xP_{\ell}(x)P_{\ell^{\prime}}(x)}{x-s}dx$ and $\Omega_{00}(s)=1+\frac{1}{2}s\ln (s-1)/(s+1)$. All distortions of order $\ell^{\prime}$= 0, 1, 2, 3,..., $\infty$ enter explicitly Eq.(10) as long as the interaction parameter $F^a_{\ell}\not=0$. 
 
We are interested in the effects of the Landau parameters $F^a_0$ and $F^a_1$ on the dispersion of the Goldstone mode and the first Silin-like ferromagnetic mode and as in the hydrodynamic model, we set $F^a_{\ell}=0$ for $\ell >1$. To calculate the dynamical response $\chi(\vec{q}, \omega)$ of the system to a transverse perturbation $\delta B^+$, $\chi(\vec{q}, \omega)=-N(0)\nu^+_0/\delta B^+$, we write Eq.(10) for $\ell=0$ and use the net magnetization conservation law, Eq.(4), to express $\nu^+_1$ in terms of $\nu^+_0$ and $\delta B^+$. The real poles of the response function $\chi(\vec{q}, \omega)$

\begin{equation}
\Omega_{00}\left(sqv_F-\frac{\Delta}{F^a_0}\right)\left[s(sqv_F+\Delta)F^a_1+
F^a_0\left(1+\frac{F^a_1}{3}\right)qv_F\right]+
(sqv_F+\Delta)\left(1+\frac{F^a_1}{3}\right)qv_F=0
\end{equation}
give the undamped spin waves. In the small $q$-limit of Eq.(11) we recover the dispersion equation from the hydrodynamic model, Eq.(7), with $F^a_2=0$.

We solve numerically Eq.(11) for $|s|>1$ and obtain two solutions $\omega^R_0(q)$ and $\omega^R_1(q)$, which we identify as the Goldstone mode and the first Silin-like ferromagnetic mode. When $q$ increases both modes $\omega^R_0(q)$ and $\omega^R_1(q)$ start to deviate from the hydrodynamic values. Close to the particle-hole continuum ($s\rightarrow \pm1$), where the higher order distortions become significant, there is a big shift in the frequency (see Fig.2 and Fig.3). To see the behavior of the modes close to the singularity we expand Eq.(11) in the small $\epsilon$-limit ($s=-1-\epsilon$ and $\epsilon\rightarrow 0^+$) and obtain

\begin{equation}
\epsilon(q)\left(1+\frac{F^a_1}{3}\right)=\left(1+\frac{\Delta}{qv_FF^a_0}\right)\left[F^a_1+\frac{F^a_0(1+F^a_1/3)}{1-\Delta/qv_F}\right].
\end{equation}
From Eq.(12) we see that there are two singularities at $q_0=-\Delta/F^a_0v_F$ and $q_1=-q_0F^a_1/(1+F^a_1/3+F^a_1/F^a_0)$. The singularity at $q_0$ is completely determined by the interaction parameter $F^a_0$ through the relation $m^0\sim|1+F^a_0|^{1/2}$ and it does not depend on the interaction parameter $F^a_1$, while the singularity at $q_1$ depends critically on $F^a_1$. The two singularities $q_0$ and $q_1$ coincide when $F^a_1=-F^a_0/(4F^a_0/3+1)$ and for $q_1>q_0$ the first Silin-like ferromagnetic mode crosses the Goldstone mode. 

\begin{figure}[h]
\includegraphics*[0.2cm,0.2cm][7.13cm,6.2cm]{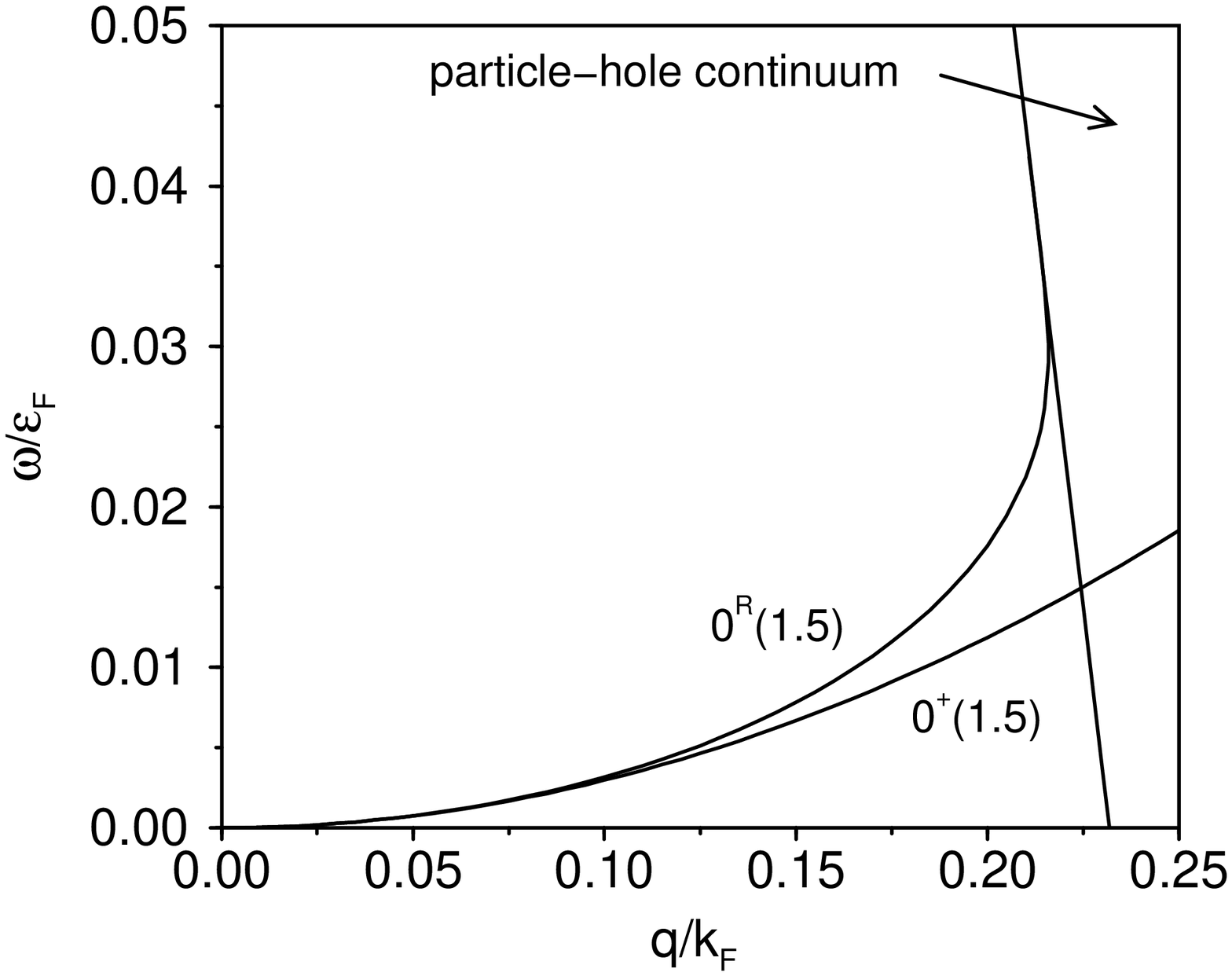}
\caption{The mode $\omega^R_0(q)\equiv 0^R$ with the hydrodynamic mode $\omega^+_0(q)\equiv 0^+$ for $F^a_0=-1.1$ and $F^a_1=1.5$.}
\end{figure}
 
We shall discuss now the behavior of the Goldstone mode in the vicinity of the singularity at $q_0$. In agreement with the RPA \cite{Ref6} the Goldstone mode touches tangentially the particle-hole continuum boundary. But for $q>q_0$ we find a branching of the solution $\omega^+_0(q)$ of Eq.(11), i.e. there are two different frequencies $\omega^+_0(q)$ for the same $q$. Close to the singularity at $q_0$, for $q\rightarrow q_0^+$ the asymptotic expansion $\omega(q)=\Delta-[1+\epsilon(q)]qv_F$ describes the behavior of the upper brunch of the Goldstone mode $\omega^R_0(q)$ (here $\epsilon\rightarrow 0^+$, $s\rightarrow-1$ and Eq.(11) has real solutions for $q\rightarrow q_0^+$). Thus our model predicts that the Goldstone mode curves upward and then touches tangentially the particle-hole continuum. It can be shown that including the higher order interaction parameter $F^a_2$ does not change the qualitative behavior of the Goldstone mode and the singularity at $q_0$ ($q_0$ does not depend on $F^a_1$ and $F^a_2$). Therefore, for $q_0>q_1$ there is no a value of the Stoner gap parameter for which the dispersion curve $\omega^R_0(q)$ goes through a maximum, as predicted by the RPA \cite{Ref6}. It is worth noting that for values of the interaction parameter $F^a_1< -F^a_0/(4F^a_0/3+1)$ the effect of crossing of the Goldstone and the first Silin-like ferromagnetic modes may result in changing the shape of the Goldstone mode in the region of larger momentum transfer, forcing it to curve down and then touch tangentially the Stoner excitation boundary, i.e. forcing the dispersion curve $\omega^R_0(q)$ to go through a maximum. 

\begin{figure}[h]
\includegraphics*[0.2cm,0.2cm][7.13cm,6.2cm]{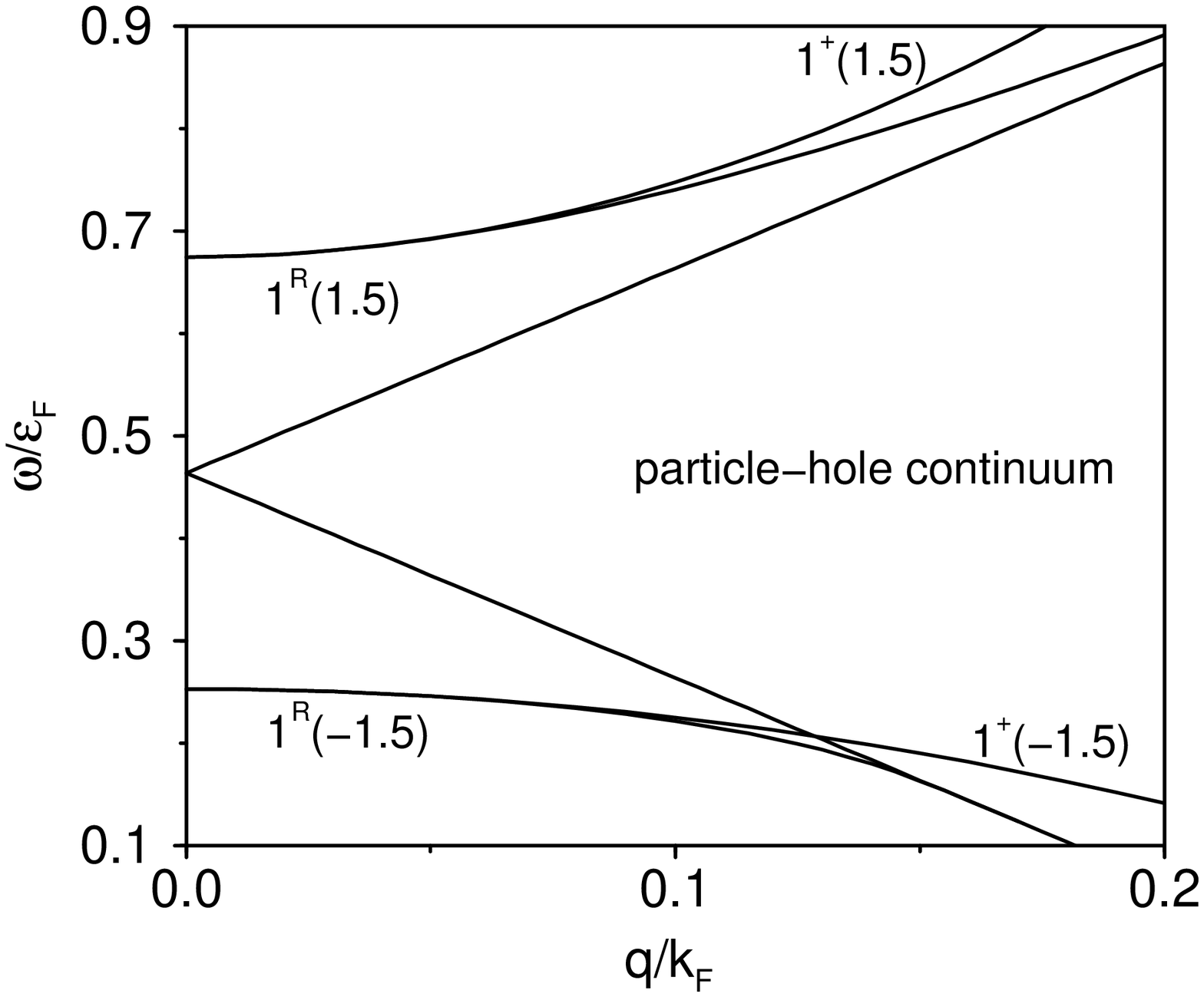}
\caption{The mode $\omega^R_1(q)\equiv 1^R$ with the hydrodynamic mode $\omega^+_1(q)\equiv 1^+$ for $F^a_0=-1.1$ and $F^a_1= -1.5, 1.5$ (shown in the brackets)}
\end{figure}

When the first Silin-like ferromagnetic mode $\omega^R_1(q)$ approaches the particle-hole continuum, it shows different behavior for different values of $F^a_1$ (see Fig.3). For negative values of $F^a_1$ the singularity at $q_1$ determines the point where $\omega^R_1(q)$ touches the particle-hole continuum, for $F^a_1=0$ the singularity is at $q_1=0$ and $\omega^R_1(q)$ degenerates to a point, for positive values of $F^a_1$ the mode $\omega^R_1(q)$ approaches asymptotically the particle-hole continuum. 

We derived above a new spin-wave mode, the first Silin-like ferromagnetic mode. The mode is propagating for any values of the interaction parameter $F^a_1\neq 0$. Now the obvious question is why the first Silin-like ferromagnetic mode has not been observed by inelastic neutron measurements on weak ferro magnets? To answer to this question, we calculate the intensities \cite{Ref14} of the hydrodynamic Goldstone and first Silin-like ferromagnetic modes, extending our model to low temperatures and assuming the regime where the spin-diffusion lifetime varies with the temperature as $\tau_D\propto1/T^2$. It can be shown that the intensity of the Goldstone mode is about 3 orders of magnitude higher than the intensity of the first Silin-like ferromagnetic mode, the linewidth $\Gamma_0$ of the Goldstone mode is much less than the linewidth $\Gamma_1$ of the first Silin-like ferromagnetic mode (the ratio $\Gamma_1/\Gamma_0=1+(1+F^a_1)\omega^+_1/\omega^+_0(q)$ depends on the Landau parameters $F^a_0$ and $F^a_1$) and with increasing the temperature the intensities of both modes decrease and broaden. 

In summary: our main results, the dispersion laws for the Goldstone mode and for the first Silin-like ferromagnetic mode, are obtained as an implication of the FFLT. Including the interaction parameter $F^a_1$ brings in new physics -- the presence of a new spin wave mode, the first Silin-like ferromagnetic mode. The framework presented in this paper allows us to obtain the whole picture for the dispersion and attenuation of the spin waves in ferromagnetic metals.

\end{document}